\RequirePackage{fix-cm}
\documentclass[smallcondensed]{svjour3}     
\smartqed  
\usepackage{graphicx}

\usepackage{amssymb}
\usepackage{amsmath}
\usepackage{subcaption}

\usepackage{amsfonts}

\usepackage{epsfig}
\usepackage{epstopdf}
\usepackage{setspace}

\usepackage{color}

\usepackage{hyperref}
\hypersetup{
    colorlinks=true,
    citecolor=red,
    linkcolor=blue,
    filecolor=magenta,      
    urlcolor=cyan}

\usepackage{color}
\usepackage[sort&compress,numbers]{natbib}

\usepackage[demo]{adjustbox}
\usepackage{xcolor}
\usepackage{atbegshi}
\AtBeginDocument{\AtBeginShipoutNext{\AtBeginShipoutDiscard}}	
\PassOptionsToPackage{hyphens}{url}\usepackage{hyperref} 

\begin{document}
	\pagestyle{empty} 
	
		\color[rgb]{.4,.4,1}
		\hspace{5mm}

		\bigskip
		
		\hspace{15mm}
		\begin{minipage}{10mm}
			\color[rgb]{.7,.7,1}
			\rule{1pt}{226mm}
		\end{minipage}
		\begin{minipage}{133mm}
			\vspace{10mm}        
			\color{black}
			\sffamily
			\LARGE\bfseries Random vibrations of stress-driven nonlocal beams with external damping  \\[-0.3\baselineskip]   \\[-0.3\baselineskip] 
			
			\vspace{50mm}
			{\large {Preprint of the article published in \\[-0.1\baselineskip] Meccanica \\
			[-0.1\baselineskip] DOI: 10.1007/s11012-020-01181-7 }} 
			
			\vspace{30mm}        
			{\large Francesco P. Pinnola,\\[-0.1\baselineskip] Marzia S. Vaccaro Barretta,\\[-0.1\baselineskip] \textsc{Raffaele Barretta}, \\[-0.1\baselineskip] Francesco Marotti de Sciarra} 
			
			\large
			
			\vspace{40mm}
			\vspace{25mm}
			
			\small
			\url{https://doi.org/10.1007/s11012-020-01181-7}
			
			\textcircled{c} 2020. This manuscript version is made available under the CC-BY-NC-ND 4.0 license \url{http://creativecommons.org/licenses/by-nc-nd/4.0/}
			\hspace{30mm} 
			\color[rgb]{.4,.4,1} 
		\end{minipage}


\title{Random vibrations of stress-driven nonlocal beams with external damping
}


\author{Francesco P. Pinnola         \and
        Marzia S. Vaccaro		\and
        Raffaele Barretta \and
        Francesco Marotti de Sciarra 
}

\institute{F.P. Pinnola \and M.S. Vaccaro \and R. Barretta \and F. Marotti de Sciarra\at
              Department of Structures for Engineering and Architecture, \\
              University of Naples Federico II, \\
via Claudio 21, Ed. 6, 80125 - Naples, Italy\\
              \email{\url{rabarret@unina.it} }
}

\date{Received: date / Accepted: date}

\maketitle

\begin{abstract}
Stochastic flexural vibrations of small-scale Bernoulli-Euler beams with external damping are investigated by stress-driven nonlocal mechanics.
Damping effects are simulated considering viscous interactions between beam and surrounding environment.
Loadings are modeled by accounting for their random nature. 
Such a dynamic problem is characterized by a stochastic partial differential equation in space and time governing time-evolution of the relevant displacement field. 
Differential eigenanalyses are performed to evaluate modal time coordinates and mode shapes, 
providing a complete stochastic description of response solutions.
Closed-form expressions of power spectral density, correlation function, stationary and non-stationary variances of displacement fields are analytically detected.  
Size-dependent dynamic behaviour is assessed in terms of stiffness, variance and power spectral density of displacements.
The outcomes can be useful for design and optimization of structural components of modern small-scale devices, such as Micro- and Nano-Electro-Mechanical-Systems (MEMS and NEMS). 

\end{abstract}

\section{Literature survey, motivation and outline}
Methodologies to predict random vibrations in structural systems have reached 
over the last century a significant importance in design and optimization of new-generation composites 
\cite{Pourasghar2019,Xia2019} and technological devices, such as: 
micro-bridges \cite{Mojahedi2017}, nano-switches \cite{Moradweysi2018}, 
nano-generators \cite{Hosseini,debellis}, nano-sensors \cite{Mohammadian,natsuki},
energy harvesters \cite{Tran2018,Basutkar2019,GhayeshFarokhi2020}.
It is acknowledged that continuum mechanics is able to model structural components of small-scale systems \cite{FarajpourReview2018}, but some mechanical expedients are needed to accurately predict nonconventional  phenomena, such as size and damping effects. 
It is well-established that inter-atomic forces and molecular interactions cannot be overlooked in
small-scale structures which exhibit technically significant size effects 
\cite{Bauer,Kiang,Xiao,Zienert,Chowdhury,Tang}.
Such a phenomenon cannot be captured by the classical theory of local continuum \cite{marB} due
to lack of internal characteristic scales. 
Nonlocal continua, driven by suitably chosen scale characteristic parameters, 
are instead appropriate to model micro- and nano-structures \cite{3,6,alo16,marA}, as confirmed by molecular dynamic simulations \cite{Ansari2010,Murmu,Ansari2012}. 
Nonlocal methodologies allow for modeling complex mechanical behaviours avoiding computationally expensive procedures \cite{1,2}. 

Nonlocal theory, in its earliest formulation \cite{Rogula1965,Rogula1982} is based on the idea that the stress at a point of a continuum depends not only on elastic strain at that point but involves local responses of the whole structure. 
Long-range interactions are thus described by a strain-driven convolution integral supplemented with an averaging kernel characterized by a nonlocal length-scale parameter. 
Such an approach was consistently applied by Eringen \cite{Eringen,Eringen1} to nonlocal problems involving screw dislocation and wave propagation, that are formulated in unbounded domains.
However, mathematical difficulties are apparent when the strain-driven strategy is applied to structural problems which are generally defined in bounded domains. 

In fact, assuming that the averaging kernel is the Green function of a differential operator, from the integral convolution it is possible to obtain a consequent differential formulation \cite{Tricomi1985,Polyanin2008}. 
For bounded structural domains, to the differential formulation above is needed to add a proper set of Constitutive Boundary Conditions (CBCs) \cite{BarMar}.
Paradoxical results and unacceptable nonlocal responses are obtained \cite{Challamel2008,ReddyParadoxSolved}
if CBCs are ignored. 
Several nonlocal theories have been adopted to overcome the aforementioned difficulties, such as: two-phase models \cite{7,KhodabakhshiReddy2015}, strain and stress gradient theories 
\cite{4,ChallamelReddy2016,CivalekIJES2018,CornacchiaJCOMB2019,CornacchiaMAMS2019}, 
nonlocal gradient techniques \cite{LimJMPS2015,BarrettaMar2018,ApuzzoBarretta2018},
strain-difference approaches \cite{48,FusPisPol,marC}, displacement-based nonlocal models \cite{DiPaola,pir1,alo13,alo15}, stress-driven formulation of nonlocality \cite{RomBar,BarVacc}. 
Advantageously, the stress-driven approach has been shown to be able to effectively model the nonlocal behaviour of small-scale structures and provides exact solutions for problems of applicative interest 
in Nano-Engineering \cite{41,BarMar2}. 
The stress-driven nonlocal integral model is therefore adopted in this paper to significantly tackle size-dependent random vibrations in inflected elastic small-scale beams. 
A first effort on the matter, disregarding random phenomena, was performed in \cite{ApuBarMar}.

Another import effect in dynamics of micro- and nano-systems concerns damping phenomena which should be properly modeled in applicative problems of nano-engineering. 
Instances are listed as follows: external magnetic force \cite{LeeLin2010}, humidity, thermal and paddling effects \cite{Chen2011} and internal viscous force due to material rheological properties \cite{DiPPinVal}. 
Modeling of internal viscous forces is successfully performed by properly selecting constitutive formulations \cite{DiMDiPPin}.
Other effects, mentioned above, are related to surrounding environmental interactions \cite{Calleja2012}. 
In the present paper, particular attention is paid to capture external damping effects, useful to analyse modern small-scale structures, such as: 
sensors inside viscous fluid, devices under magnetic field, nano-systems for biological detection. 
Specifically, a bed of independent dashpots will be considered to simulate external viscous interactions between nonlocal beam and surrounding environment.

At micro- and nano-scales, structures can be excited by different kind of force  systems. 
An example is the effect of environmental thermal and/or mechanical noises in nano-sensors \cite{Mems}. 
Stochastic approaches can be conveniently exploited to model external loadings \cite{Verma,Spanos,Crandal}, effectively representable by random time-process \cite{Pirr1,Pirr2,AloDiPPin}.

For the aforesaid reasons, the present research provides a novel strategy for stress-driven nonlocal analysis of damped vibrations of elastic nano-beams due to stochastic excitation. 
Steady-state solutions are established, detecting thus analytical expressions of power spectral density and stationary variance of displacements. 
Closed-form solutions are also evaluated for non-stationary responses of nonlocal damped beams forced by Gaussian white noise.


The manuscript is organized as follows. 
Strain- and stress-driven models of pure nonlocal integral elasticity are recalled and specialized to Bernoulli--Euler beams in Section~\ref{sect1}.
Dynamic equilibrium equations of damped beams are established in Section~\ref{DM} by using the well-posed stress-driven nonlocal strategy of elasticity.
Mode shape functions and natural frequencies are analytically detected  
and an effective methodology to perform dynamic eigenanalysis is also elucidated. 
A stochastic analysis of nonlocal damped beams forced by Gaussian white noise is developed
in Section~\ref{sect4}. 
Both stationary and non-stationary examinations are performed, detecting thus closed form solutions of displacement variance, power spectral density and correlation function. 
Numerical simulations are implemented in Section~\ref{sect5} to test accuracy of obtained solutions.
Analytical stationary and non-stationary variances are compared with numerical outcomes derived by Monte Carlo simulations. 
A parametric study of stochastic responses, in terms of displacement variances and natural frequencies, is given in Section~\ref{sect5} to study nonlocal effects.
Closing remarks are outlined in Section~\ref{sect6}.

\section{Purely nonlocal integral elasticity}
\label{sect1}

Two purely nonlocal models of elasticity are available in literature:
\begin{enumerate}
\item
strain-driven integral law \cite{Rogula1965}, applied to problems of screw dislocation and wave propagation \cite{Eringen,Eringen1};
\item
stress-driven integral law \cite{RomBar} applied to nonlocal mechanics of structures.
\end{enumerate}
These theories are preliminarily recalled below for $\,3$-D continua and specialized to Bernoulli-Euler beams.
Eringen's strain-driven law is based on the idea that the stress $\boldsymbol{\sigma}$ at a point $\boldsymbol{x}$ of a nonlocal $\,3$-D continuous body $\mathcal{B}$ is the output of a convolution between the local response to the elastic strain field $\boldsymbol{\varepsilon}$ and a scalar kernel $\Phi_\lambda$ depending on a non-dimensional positive nonlocal parameter $\lambda$. That is,
\begin{equation}
\boldsymbol{\sigma}(\boldsymbol{x})=\int_\mathcal{B}\Phi_\lambda(\boldsymbol{x},\bar{\boldsymbol{x}})\boldsymbol{E}(\bar{\boldsymbol{x}})\,\boldsymbol{\varepsilon}(\bar{\boldsymbol{x}})d\bar{\boldsymbol{x}}
\label{eq2.01}
\end{equation}
with $\boldsymbol{E}$ fourth-order local elasticity stiffness tensor. 

For a Bernoulli-Euler inflected beam of length $\,L\,$, the nonlocal strain-driven relation Eq.~(\ref{eq2.01})
takes the form
\begin{equation}
M (z)= EI\int_0^L\Phi_\lambda(z,\bar z)\chi (\bar z)d\bar z
\label{eq2.2}
\end{equation}
with 
$\,M\,$ bending moment,
$\,E\,$ Euler-Young modulus, $\,I\,$ cross-sectional moment of inertia along the bending axis $\,y\,$,
$\,z\,$ beam axial abscissa and $\,\chi\,$ elastic curvature. 
The integral kernel $\Phi_\lambda$ can be selected among exponential, Gaussian or power-law type functions and must satisfy properties of positivity, symmetry and limit impulsivity \cite{Eringen1}. 
A convenient choice for the averaging kernel is the special bi-exponential function 
\begin{equation}
\Phi_\lambda(z,\bar z)=\frac{1}{2 \lambda L}\exp\left({-\frac{|z-\bar{z}|}{\lambda L}}\right)
\label{eq2.3}
\end{equation}
where $\lambda L$ is the characteristic length $L_c$. 
With the assumption above, 
the integral law Eq.~\eqref{eq2.2} is equivalent to the second-order differential equation \cite{BarMar}
\begin{equation}
M^{(2)}(z)-\frac{1}{(\lambda L)^2}M (z)=-\frac{EI }{(\lambda L)^2}\chi (z)
\label{eq2.4}
\end{equation}
supplemented with the following constitutive boundary conditions (CBC)
\begin{equation}
\left\{\begin{split}
&M^{(1)}(0)=\frac{1}{\lambda L}M (0)\\
&M^{(1)}(L)=-\frac{1}{\lambda L}M (L)
\end{split}\right.
\label{eq2.5}
\end{equation}

It is worth noting that, for structural problems of applicative interest, CBCs Eq.~\eqref{eq2.5} are in contrast with 
equilibrium requirements \cite{BarMar}.
Incompatibility between equilibrium and constitutive conditions reveals that Eringen's nonlocal model leads to ill-posed structural problems, generally formulated in bounded domains. 
Such an obstruction can be overcome by using the stress-driven approach \cite{RomBar} which is well-posed.  
Exact nonlocal structural solutions can be found in \cite{BarMar2}.  

In stress-driven mechanics, the elastic strain $\boldsymbol{\varepsilon}$ at a point $\boldsymbol{x}$ of $\mathcal{B}$ is obtained by convoluting the local stress $\boldsymbol{\sigma}$ with an averaging kernel
$\,\Phi_\lambda\,$
\begin{equation}
\boldsymbol{\varepsilon}(\boldsymbol{x})=\int_\mathcal{B}\Phi_\lambda(\boldsymbol{x},\bar{\boldsymbol{x}})\boldsymbol{C}(\bar{\boldsymbol{x}})\,\boldsymbol{\sigma}(\bar{\boldsymbol{x}})d\bar{\boldsymbol{x}}
\label{eq3.1a}
\end{equation}
with $\boldsymbol{C}=\boldsymbol{E}^{-1}$ local elastic compliance. 
The stress-driven model for Bernoulli-Euler beams is governed by the following moment-curvature relation
\begin{equation}
\chi (z)=\frac{1}{EI}\int_0^L \Phi_\lambda(z-\bar z)M (\bar{z})d\bar z 
\label{eq3.0}
\end{equation}

The integral formulation Eq.~\eqref{eq3.0}, with the special kernel Eq.~\eqref{eq2.3}, 
is equivalent to the second order differential equation \cite{RomBar}
\begin{equation}
\chi^{(2)}(z)-\frac{1}{(\lambda L)^2}\chi(z)=-\frac{1}{EI (\lambda L)^2}M ({z}) 
\label{eq3.01}
\end{equation}
equipped with the following constitutive boundary conditions 
\begin{equation}
\left\{\begin{split}
&\chi^{(1)}(0)=\frac{1}{\lambda L}\chi (0)\\
&\chi^{(1)}(L)=-\frac{1}{\lambda L}\chi (L)
\end{split}\right.
\label{eq3.02}
\end{equation}

The stress-driven approach Eqs.~\eqref{eq3.01}, \eqref{eq3.02} provides exact solutions in both static and dynamic structural problems and is exploited in the present study to analytically tackle size-dependent random vibrations of slender elastic beams.

\section{Dynamical analysis of nonlocal beams}
\label{DM}

\begin{figure}[h]
  \begin{center}
        \includegraphics[width=0.65\textwidth]{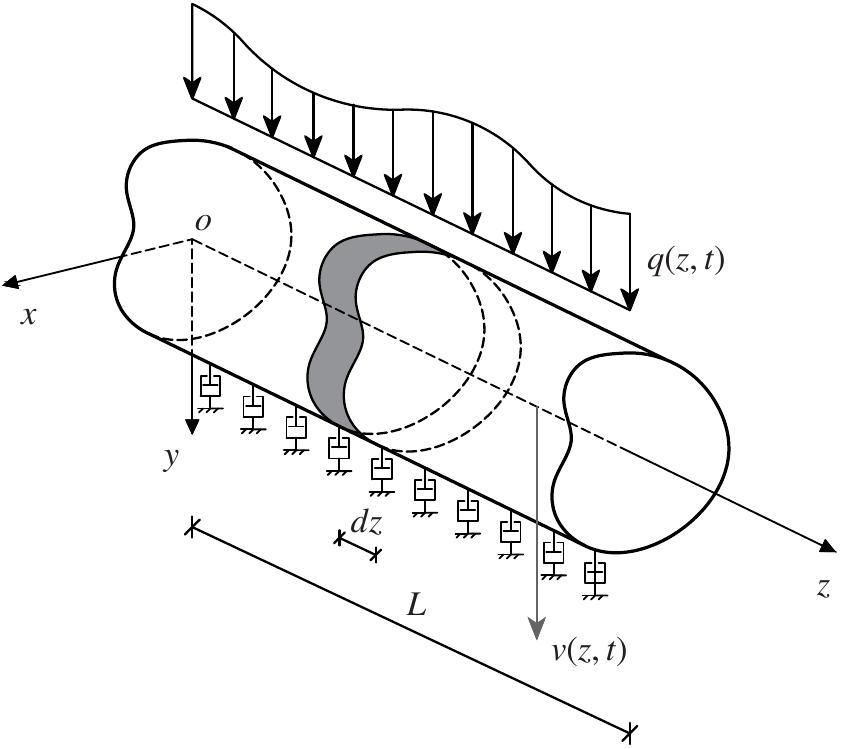}
    \caption{Bernoulli-Euler beam with external damping}
    \label{fig1a}
  \end{center}
\end{figure}

Let us consider a nonlocal beam of length $L$ and cross-sectionial area $A_t$
subjected to a transverse loading per unit of length $q(z,t)$, see \figurename~\ref{fig1a}.
$(x, y, z)$ is the adopted system of orthogonal coordinates,
$v(z,t)$ denotes the transverse displacement and $\rho(z)$ is the mass density.
\begin{figure}[h]
\centering
\hspace{10mm}
           \includegraphics[width=0.5\textwidth]{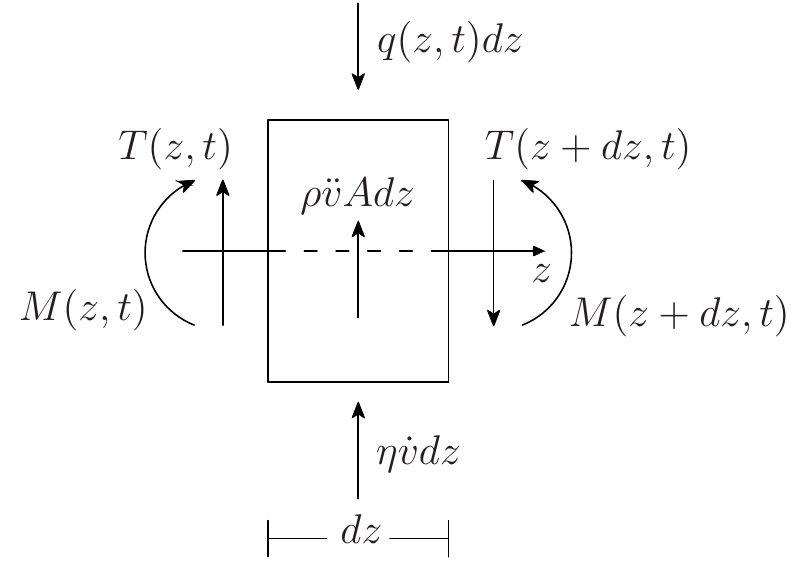}
    \caption{Free-body diagram of a beam differential element}
    \label{fig1b}
\end{figure}
The stress-driven formulation Eqs.~\eqref{eq3.01}, \eqref{eq3.02} is used as nonlocal elasticy law while the effect of external damping is introduced and modeled as a bed of dashpots with viscosity $\eta$. 
This kind of viscous interaction is able to model a possible external interaction between nonlocal beam and viscous fluid. 
Damping can be also simulated as a material effect \cite{adik,adik2,Failla,alo18,PirDim,AlottaA}
by a viscoelastic law. 

\break
In this paper, following the Newtonian approach, we consider a beam differential element of length $dz$, whose free-body diagram is represented in \figurename~\ref{fig1b}. 
The equilibrium equation along the $y$-direction, involving external loading, inertial and damping forces, and bending and shearing fields, writes as
\begin{equation}
\frac{\partial T(z,t)}{\partial z}dz+q(z,t)dz =\eta dz \frac{\partial v(z,t)}{\partial t}+\rho A dz \frac{\partial^2 v(z,t)}{\partial t^2},\;\;\;0<z<L
\label{taglio}
\end{equation}
%

By ignoring second-order terms in $dz$ and neglecting mass moment of inertia and angular acceleration, the rotational equilibrium along the $x$-axis gives
\begin{equation}
\frac{\partial M(z,t)}{\partial z}-T(z,t) =0,\;\;\;\;\;0<z<L
\label{momento}
\end{equation}

Combining the equilibrium equations along $x$- and $y$-directions we get the partial differential equation \begin{equation}
-\frac{\partial^2 M(z,t)}{\partial z^2}+\eta \frac{\partial v(z,t)}{\partial t}+\rho A \frac{\partial^2 v(z,t)}{\partial t^2}=q(z,t),\;\;\;0<z<L
\label{equilibrio}
\end{equation}
By introducing the nonlocal stress-driven relation in Eq.~(\ref{eq3.01}), from Eq.~(\ref{equilibrio}) we get
\begin{equation}
-EI \left[\frac{\partial^2\chi(z,t)}{\partial z^2}-(\lambda L)^2\frac{\partial^4\chi(z,t)}{\partial z^4}\right]+\eta \frac{\partial v(z,t)}{\partial t}+\rho A \frac{\partial^2 v(z,t)}{\partial t^2}=q(z,t)
\label{equilibrio chi}
\end{equation}

According to Bernoulli-Euler kinematics, curvature $\chi(z,t)$ is related to the transverse displacement $v(z,t)$ by 
\begin{equation}
\chi(z,t)=-\frac{\partial^2v(z,t)}{\partial z^2}
\label{chi}
\end{equation}
and by placing Eq.~(\ref{chi}) into Eq.~(\ref{equilibrio chi}) we have that
\begin{equation}
\frac{\partial^4v(z,t)}{\partial z^4}-(\lambda L)^2\frac{\partial^6v(z,t)}{\partial z^6}+\frac{\eta}{EI} \frac{\partial v(z,t)}{\partial t}+\frac{\rho A}{EI} \frac{\partial^2 v(z,t)}{\partial t^2}=\frac{q(z,t)}{EI}
\label{equilibrio EB}
\end{equation}
which is the partial differential equilibrium equation ruling bending vibrations of a nonlocal Bernoulli-Euler beam resting on a bed of dashpots. 

For vanishing nonlocal parameter $\lambda\to0^+$, Eq.~(\ref{equilibrio EB}) provides the known differential equation governing forced vibrations of local Bernoulli-Euler beams resting on a bed of independent dashpots. Moreover, setting $\eta=0$, the classical formulation of undamped local beams is obtained \cite{Meirovi,Pirrotta}. 
Solutions of the introduced partial differential equation may be found by imposing two initial conditions, four standard BCs and two constitutive BCs of the stress-driven model Eq.~(\ref{eq3.02}).

\subsection{Free vibrations of undamped nonlocal beam}
In order to solve Eq.~(\ref{equilibrio EB}), we first consider the undamped nonlocal beam in free vibration. In other words, we set $q(z,t)=0$ and $\eta=0$. These assumptions imply that Eq.~(\ref{equilibrio EB}) yields
\begin{equation}
\frac{\partial^4v(z,t)}{\partial z^4}-(\lambda L)^2\frac{\partial^6v(z,t)}{\partial z^6}+\frac{\rho A}{EI} \frac{\partial^2 v(z,t)}{\partial t^2}=0,\;\;\;0<z<L
\label{equilibrio free}
\end{equation}
which rules free vibrations of undamped nonlocal beam. We suppose that the displacement function $v(z,t)$ is with separable variables and so it can be expressed by a product of a space function $\phi(z)$ (mode shape) and a time-dependent function $y(t)$ that modulates the amplitude of mode shape in time. That is,
\begin{equation}
v(z,t)=\phi(z)y(t)
\label{variab sep}
\end{equation}
by substituting Eq.~(\ref{variab sep}) into Eq.~(\ref{equilibrio free}) we get
\begin{equation}
y(t)\left[\frac{d^4\phi(z)}{d z^4}-(\lambda L)^2\frac{d^6\phi(z)}{d z^6}\right]+\frac{\rho A}{EI}  \phi(z)\frac{d^2 y(t)}{d t^2}=0,\;\;\;0<z<L
\label{diff eigen}
\end{equation}
where partial derivatives have been replaced by total derivatives due to the assumption in Eq.~(\ref{variab sep}). Moreover, by using Lagrange's differential notation for space derivative and Newton's notation for time derivative we can rewrite Eq.~(\ref{diff eigen}) as
\begin{equation}
\frac{EI}{\rho A}\frac{(\lambda L)^2\phi^{(6)}(z)-\phi^{(4)}(z)}{\phi(z)}=\frac{\ddot y(t)}{y(t)} ,\;\;\;\;\;0<z<L
\label{diff new lagr}
\end{equation}
where both sides of Eq.~(\ref{diff new lagr}) must be equal to a constant $\alpha$ that can be associated with the natural frequency of the oscillation $\omega_0$ as 
\begin{equation}
\alpha=-\omega_{0}^2
\end{equation}

Hence,  Eq.~(\ref{diff new lagr}) can be rewritten as 
\begin{equation}
\frac{EI}{\rho A}\frac{(\lambda L)^2\phi^{(6)}(z)-\phi^{(4)}(z)}{\phi(z)}=\frac{\ddot y(t)}{y(t)} =-\omega_{0}^2
\label{eigenvalues}
\end{equation}
by considering the left side of Eq.~(\ref{eigenvalues}), we get the following sixth order differential equation in the space variable $z$ 
\begin{equation}
{(\lambda L)^2\phi^{(6)}(z)-\phi^{(4)}(z)}+\omega_{0}^2 \frac{\rho A}{EI} {\phi(z)}=0,\;\;\;\;\;0<z<L
\label{eigenfunction}
\end{equation}
whose solution must satisfy the four BCs depending on the type of loads and constraints at the bounds and the two constitutive BCs in Eq.~(\ref{eq3.02}). That is,
\begin{equation}
\left\{\begin{split}
&\phi^{(3)}(0)=\frac{1}{\lambda L}\phi^{(2)}(0)\\
&\phi^{(3)}(L)=-\frac{1}{\lambda L}\phi^{(2)} (L)
\end{split}\right.
\label{2bc}
\end{equation}

The problem to find the constant $\omega_0$ and $\phi(z)$ such that Eq.~(\ref{eigenfunction}) admits nontrivial solution is known as differential eigenvalue-eigenfunction problem, where $\omega_0$ is the eigenvalue and $\phi(z)$ represents the corresponding eigenfunction. Eq.~(\ref{eigenfunction}) admits infinite eigenvalues and then infinite eigenfunctions. Therefore, the solution in terms of displacement can be expressed as a sum of infinite products between the modal time coordinates $y_j(t)$ and the mode shapes $\phi_j(z)$
\begin{equation}
v(z,t)=\sum_{j=1}^\infty\phi_j(z)y_j(t)
\label{exact}
\end{equation}
where the $j$-th eigenfunction $\phi_j(z)$ is a solution of the $j$-th homogeneous sixth-order differential equation in Eq.~(\ref{eigenfunction}). Specifically,
\begin{equation}
\phi_j(z)=\sum_{i=1}^3 \left\{C_i \exp\left[{\sqrt{\gamma_i(\omega_{0,j})} z}\right]+C_{i+3} \exp\left[-{\sqrt{\gamma_i(\omega_{0,j})} z}\right]\right\}
\label{phi}
\end{equation}
where $\gamma_i(\omega_{0,j})$ with $i=1,2,3$ are the roots of the following characteristic third degree polynomial
\begin{equation}
(\lambda L)^2\gamma^3-\gamma^2+\omega_{0,j}^2 \frac{\rho A}{EI}=0
\label{poli}
\end{equation}
and the coefficients $C_i$ are obtained by imposing that solution in Eq.~(\ref{phi}) satisfies the previous six BCs and the normality condition. 
In this manner, each eigenfunction possesses the following orthonormality property
\begin{equation}
\int_0^L\phi_i(z)\phi_j(z) dz=\delta_{kj}
\label{orto}
\end{equation}
where $\delta_{kj}$ indicates the Kronecker delta defined as
\begin{equation}
\delta_{kj}=\left\{
\begin{split}
&1 \;\;\textrm{if}\;k=j,\\
&0 \;\;\textrm{if}\;k\neq j
\end{split}
\right.
\end{equation}

Such solution of differential eigenvalue problem will be used in the next section to solve the more general case of a forced nonlocal damped beam.

\subsection{Forced vibrations of damped nonlocal beam}
Now, we consider the dynamical problem of a forced beam with external damping. The solution in terms of displacements expressed in Eq.~(\ref{exact}) in function of mode shapes $\phi_j(z)$ is placed into the equilibrium equation~(\ref{equilibrio EB}). Specifically,
\begin{equation}
\begin{split}
&\frac{EI}{\rho A}\left[\sum_{j=1}^\infty y_j(t)\phi^{(4)}_j(z)-\sum_{j=1}^\infty y_j(t)(\lambda L)^2\phi^{(6)}_j(z)\right]+     \\
&+\frac{\eta}{\rho A}\sum_{j=1}^\infty \dot y_j(t)\phi_j(z)+ \sum_{j=1}^\infty \ddot y_j(t)\phi_j(z)=\frac{q(z,t)}{\rho A}
\label{eq forc}
\end{split}
\end{equation}
by multiplying both sides of Eq.~(\ref{eq forc}) by the \textit{i}-th eigenfunction $\phi_i(z)$ and integrating on the domain $[0,L]$, Eq.~(\ref{eq forc}), we get
\begin{equation}
\begin{split}
&\frac{EI}{\rho A}\left[\sum_{j=1}^\infty y_j(t)  \int_0^L\phi_i(z)\phi^{(4)}_j(z)dz\right]+  \\
&-\frac{EI}{\rho A}\left[\sum_{j=1}^\infty y_j(t)(\lambda L)^2  \int_0^L\phi_i(z)\phi^{(6)}_j(z)dz\right]+\\
&+\frac{\eta}{\rho A} \sum_{j=1}^\infty \dot y_j(t) \int_0^L\phi_i(z)\phi_j(z)dz+ \sum_{j=1}^\infty \ddot y_j(t) \int_0^L\phi_i(z)\phi_j(z)dz=  \\
&\frac{1}{\rho A}\int_0^L\phi_i(z)q(z,t)dz
\label{eq forc 2}
\end{split}
\end{equation}

Taking into account the eigenfunctions orthonormality property in Eq.~(\ref{orto}), from Eq.~(\ref{eq forc 2}) we get the second order differential equation in terms of modal coordinate $y_i(t)$ that rules the motion of a forced damped modal oscillator
\begin{equation}
\frac{EI}{\rho A}y_i(t)\left[a_i-(\lambda L)^2 b_i\right]+\frac{\eta}{\rho A} \dot y_i(t)+  \ddot y_i(t) =\frac{1}{\rho A}\int_0^L\phi_i(z)q(z,t)dz
\label{eq forc 3}
\end{equation}
where the coefficients $a_i$ and $b_i$ are
\begin{equation}
a_i=\int_0^L\phi_i(z)\phi^{(4)}_i(z)dz,\;\;\;\;\;
b_i=\int_0^L\phi_i(z)\phi^{(6)}_i(z)dz
\label{ak bk}
\end{equation}

Moreover, we assume that also the load is with separable variables. Therefore,
\begin{equation}
q(z,t)=g(z)f(t)
\label{load}
\end{equation}
this implies that Eq.~(\ref{eq forc 3}) can be rewritten as
\begin{equation}
\frac{k_{\lambda, i}}{\rho A}y_i(t)+\frac{\eta}{\rho A} \dot y_i(t)+  \ddot y_i(t) =\frac{c_i}{\rho A} f(t)
\label{eq forc 4}
\end{equation}
where the \emph{nonlocal modal stiffness} $k_{\lambda,i}$ is defined as
\begin{equation}
{k}_{\lambda, i}=EI\left[a_i-(\lambda L)^2b_i\right]
\end{equation}
and the coefficient $c_i$ is 
\begin{equation}
c_i=\int_0^L\phi_i(z)g(z)dz
\end{equation}

Notice that the following term
\begin{equation}
\omega_{0,i}=\sqrt{\frac{k_{\lambda, i}}{\rho A}}
\end{equation}
represents the natural frequency associated to the $i$-th mode $\phi_j(z)$.

Solution of Eq.~(\ref{eq forc 4}) provides the \textit{i}-th modal time coordinate of the infinite series in Eq.~(\ref{exact}).
Truncating the sum to an appropriate number $n$ of eigenfunctions and modal coordinates leads to the following approximated solution, that will be used for the numerical applications
\begin{equation}
v(z,t)\approx\sum_{j=1}^n\phi_j(z)y_j(t)
\label{approx}
\end{equation}

\section{Stochastic analysis of nonlocal beams}\label{sect4}
Now we turn the attention to random vibrations of nonlocal beams, supposing that the time-dependent part of the transverse load $q(z,t)$ has a stochastic nature. Specifically, from Eq.~(\ref{load}) we assume that
\begin{equation}
q(z,t)=g(z)F(t)
\label{load2}
\end{equation}
where $g(z)$ is a deterministic function, while $F(t)$ is a stochastic process, the latter is noted by a capital letter to distinguish it from the deterministic function $f(t)$. Moreover, we assume that $F(t)$ is a stationary Gaussian process with zero mean $\mu_F$ and with an assigned correlation function (CF) denoted by $R_F(\tau)$. 
Thus, the input process is completely described by the following time-independent parameters
	\begin{subequations}
\begin{equation}
\mu_F=\mu_F(t):=\mathbb{E}\left[F(t)\right]=0
\label{mean F}
\end{equation}
\begin{equation}
\begin{split}
R_F(\tau)=R_F(t,t+\tau):&=\mathbb{E}\left[F(t)F(t+\tau)\right]-\mu_F^2\\
&=\mathbb{E}\left[F(t)F(t+\tau)\right]
\end{split}
\label{cf}
\end{equation}
	\end{subequations}
where $\mathbb{E}[\cdot]$ is the averaging operator. For $\tau=0$ the CF gives the value of the variance $\sigma_F^2$. That is,
\begin{equation}
\sigma_F^2=\mathbb{E}\left[F(t)F(t)\right]=R_F(0)
\end{equation}

Being $F(t)$ a stationary process, by virtue of Wiener-Khinchin theorem, the power spectral density (PSD), denoted by $S_F(\omega)$, is the Fourier transform of the correlation function. That is,
\begin{equation}
\begin{split}
S_F(\omega):&=\frac{1}{2\pi}\int_{-\infty}^\infty R_f(\tau)e^{-\mathrm{i}\omega \tau}d\tau\\
			&=\lim_{\mathrm{T}\to\infty}\frac{1}{2\pi \mathrm{T}}\mathbb{E}\left[\hat F^*(\omega,\mathrm{T})\hat F(\omega,\mathrm{T})\right]
\end{split}
\label{psd f}
\end{equation}
where $\mathrm{i}=\sqrt{-1}$ is the imaginary unit, $\hat{F}(\omega,\mathrm{T})$ indicates the truncated Fourier transform of the process $F(t)$ in a finite time interval $[0,\mathrm{T}]$, and $\hat{F}^*(\omega,\mathrm{T})$ denotes its complex conjugate.

By taking into account Eq.~(\ref{load2}) and the definition in Eq.~(\ref{cf}), CF of the loading $q(z,t)$ is given by
\begin{equation}
\begin{split}
R_q(z,\tau)	&=\mathbb{E}\left[q(z,t)q(z,t+\tau)\right]\\
			&=g(z)\mathbb{E}\left[F(t)F(t+\tau)\right]g(z)\\
			&=g^2(z)R_F(\tau)
\end{split}
\label{}
\end{equation}
Similarly, the PSD of the loading is
\begin{equation}
\begin{split}
S_q(z,\omega)&=\lim_{\mathrm{T}\to\infty}\frac{1}{2\pi \mathrm{T}}\mathbb{E}\left[\hat q^*(z,\omega,\mathrm{T})\hat q(z,\omega,\mathrm{T})\right]\\
&=g(z)\lim_{\mathrm{T}\to\infty}\frac{1}{2\pi \mathrm{T}}\mathbb{E}\left[\hat F^*(\omega,\mathrm{T})\hat F(\omega,\mathrm{T})\right]g(z)\\
&=g^2(z)S_F(\omega)
\end{split}
\label{}
\end{equation}
and the variance $\sigma_q^2(z)$ is 
\begin{equation}
\sigma^2_q(z)	=\mathbb{E}\left[q(z,t)q(z,t)\right]=g(z)\sigma^2_Fg(z)
\end{equation}

Now we want to characterize the response process in terms of displacement $v(z,t)$ when the input is a Gaussian stationary process. Being the stochastic input Gaussian also the stochastic output will be Gaussian, but the response process will have a stationary part  for $t\gg0$ and a transient non-stationary one. Therefore, it is needed the evaluation of the evolution in time of the statistics.

\subsection{Time-domain response}
By observing Eq.~(\ref{eq forc 4}) we deduce that if the forcing load $f(t)=F(t)$, then also the time response of the beam in terms of modal coordinate will be a stochastic process $Y(t)$. Hence, from Eq.~(\ref{exact}) we get
\begin{equation}
v(z,t)=\sum_{j=1}^\infty\phi_j(z)Y_j(t)
\label{comp modal}
\end{equation}
where $\phi_j(z)$ is a deterministic function that can be analytically evaluated. While, the stochastic response $Y_j(t)$ is solution of the following stochastic differential equation
\begin{equation}
  \ddot Y_j(t) +\frac{\eta}{\rho A} \dot Y_j(t)+\frac{k_{\lambda,j}}{\rho A}Y_j(t)=\frac{c_j}{\rho A} F(t)
\label{eq forc sto}
\end{equation}

The forced input $F(t)$ is a Gaussian process and the differential equation in Eq.~(\ref{eq forc sto}) is linear. This implies that the output process $Y_j(t)$ will be Gaussian too, and then can be completely characterized by the mean $\mu_{Y_j}(t)$ and the correlation function $R_{Y_j}(t,t+\tau)$. Under the assumption that the beam is quiescent at $t=0$, the two initial conditions become
\begin{equation}
\begin{cases}
v(z,0)=0,\,\forall z\in [0,L]\Rightarrow Y_j(0)=0,\, \forall j\in\mathbb{N}^+\\
\dot v(z,0)=0,\,\forall z\in [0,L]\Rightarrow \dot Y_j(0)=0,\, \forall j\in\mathbb{N}^+\\
\end{cases}
\end{equation}
The output process is obtained by applying the Duhamel superposition integral 
\begin{equation}
Y_j(t)=\frac{c_j}{\rho A}\int_0^th_j(t-\tau)F(\tau)d\tau
\label{duhamel}
\end{equation}
being $h_j(t)$ a deterministic function which represents the impulse response of the $j$-th modal oscillator. 
Such a function is defined as
\begin{equation}
h_j(t)=\sqrt{\frac{4(\rho A)^2}{4k_{\lambda,j\rho A}-\eta^2}}\exp\left(-\frac{\eta}{2\rho A}t\right)\sin\left(\sqrt{\frac{4k_{\lambda,j}\rho A-\eta^2}{4(\rho A)^2}}t\right)
\end{equation}
We can observe that the term
\begin{equation}
\omega_{D,j}=\sqrt{\frac{4k_{\lambda,j}\rho A-\eta^2}{4(\rho A)^2}}
\end{equation}
is the damped frequency. Therefore, the impulse response can be rewritten as
\begin{equation}
h_j(t)=\frac{1}{\omega_{D,j}}\exp\left(-\frac{\eta}{2\rho A}t\right)\sin\left(\omega_{D,j}t\right)
\end{equation}

Taking Eq.~(\ref{mean F}) into account and by applying the averaging operator to Eq.~(\ref{duhamel}) we can prove that the mean of the response process is zero. That is,
\begin{equation}
\begin{split}
\mu_{Y_j}(t)&=\mathbb{E}\left[\frac{c_j}{\rho A}\int_0^th_j(t-\tau)F(\tau)d\tau\right]\\
&=\frac{c_j}{\rho A}\int_0^th_j(t-\tau)\mathbb{E}	\left[F(\tau)\right]d\tau\\
&=\frac{c_j}{\rho A}\int_0^th_j(t-\tau)\mu_F(\tau)d\tau=0
\end{split}
\label{mean j}
\end{equation}
Eq.~(\ref{mean j}) implies that also the mean of displacements is zero $\mu_v(z,t)=0$ for all $t\geqslant0$, and for $z\in[0,L]$.

CF of the response process $Y_j(t)$ can be evaluated taking into account the assumption in Eq.~(\ref{cf}) and the Duhamel integral in Eq.~(\ref{duhamel}). We apply the averaging operator to the process $Y_j(t)$ considering two different time-step $t=t_1$ and $t+\tau=t_2$. That is,
\begin{equation}
\begin{split}
R_{Y_j}(t_1,t_2):=&\mathbb{E}\left[Y_j(t_1)Y_j(t_2)\right]\\
		=&   \frac{c_j^2}{(\rho A)^2}\int_0^{t_1}\int_0^{t_2} h_j(t_1-\tau_1)h_j(t_2-\tau_2)\mathbb{E}\left[F(\tau_1)F(\tau_2)  \right] d\tau_1d\tau_2     \\
		=& \frac{c_j^2}{(\rho A)^2}\int_0^{t_1}\int_0^{t_2} h_j(t_1-\tau_1)h_j(t_2-\tau_2)R_F(\tau_2-\tau_1)d\tau_1d\tau_2      
\end{split}
\label{cf y}
\end{equation}
from Eq.~(\ref{cf y}) the variance $\sigma^2_{Y_j}(t)$ can be also evaluated by placing $t_1=t_2$. 

By taking into account the Eq.~(\ref{comp modal}) the CF of the displacement $v(z,t)$ is
\begin{equation}
\begin{split}
R_v(z,t_1,t_2):&=\mathbb{E}\left[v(z,t_1)v(z,t_2)\right]\\
	&=\sum_{j=1}^\infty\sum_{i=1}^\infty\phi_j(z)R_{Y_jY_i}(t_1,t_2)\phi_i(z)
\end{split}
\label{comp modal R}
\end{equation}
where $R_{Y_jY_i}(t_1,t_2)$ is the cross correlation of the modal response processes $Y_j(t$) and $Y_i(t)$ defined as
\begin{equation}
\begin{split}
R_{Y_jY_i}(t_1,t_2):=&\mathbb{E}\left[Y_j(t_1)Y_i(t_2)\right]\\
		=& \frac{c_jc_i}{(\rho A)^2}\int_0^{t_1}\int_0^{t_2} h_j(t_1-\tau_1)h_i(t_2-\tau_2)R_F(\tau_2-\tau_1)d\tau_1d\tau_2      
\end{split}
\label{cf yji}
\end{equation}

If $t_1=t_2=t$ Eq.~(\ref{comp modal R}) provides the non-stationary variance of the displacement $v(z,t)$. That is,
\begin{equation}
\begin{split}
\sigma^2_v(z,t):&=\mathbb{E}\left[v(z,t)v(z,t)\right]\\
	&=\sum_{j=1}^\infty\sum_{i=1}^\infty\phi_j(z)\sigma^2_{Y_jY_i}(t)\phi_i(z)
\end{split}
\label{comp modal var}
\end{equation}
where $\sigma^2_{Y_jY_i}(t)$ is 
\begin{equation}
\begin{split}
\sigma^2_{Y_jY_i}(t):=&\mathbb{E}\left[Y_j(t)Y_i(t)\right]\\
		=& \frac{c_jc_i\sigma^2_F}{(\rho A)^2}\int_0^{t}\int_0^{t} h_j(t-\tau)h_i(t-\tau)d\tau d\tau\\
		=&\frac{c_jc_i\sigma^2_F}{(\rho A)^2}t\int_0^{t} h_j(t-\tau)h_i(t-\tau)d\tau        
\end{split}
\label{cf yji var}
\end{equation}
and represents the cross variance of the modal response processes $Y_j(t$) and $Y_i(t)$.

\subsubsection{Monte Carlo simulation}\label{MC}
In some cases, Eq.~(\ref{cf yji}) and Eq.~(\ref{cf yji var}) cannot be evaluated in closed form and it is needed a numerical approach to characterized the response process form a stochastic point of view. In this context,  
Monte Carlo (MC) method is a powerful tool that provides a time-domain response with the aid of digital simulations. Specifically, as a first step, it is needed the generation of a proper number $N$ of samples (or realizations) of the stochastic input $F(t)$. The $i$-th realization of the stochastic input process is denoted as $F^i(t)$ and can be generated by harmonic superposition method proposed by Shinozuka and Deodatis \cite{Shino}. According to this approach the generic $i$-th sample of the forced process is given as
\begin{equation}
F^i(t)=\sqrt{2}\sum_{j=1}^m\sqrt{2 S_F(\omega_j)\Delta\omega}\cos{\left( \omega_j t+\theta_j^i\right)}
\label{MC F}
\end{equation}
where $\omega_j=j\Delta\omega$, $\Delta\omega$ is the discretization step in the frequency domain of the PSD function $S_F(\omega)$, $\theta_j^i$ represents $i$-th realization of the independent random phase with uniform distributed probability density function between $0$ and $2\pi$. 

As second step, it is needed to evaluate the response samples. In particular, for each input sample $F^i(t)$ we need to evaluate the output process $Y^i_j(t)$ in Eq.~(\ref{eq forc sto}) with the aid of the Duhamel superposition integral in Eq.~(\ref{duhamel}). Thus,
\begin{equation}
Y_j^i(t)=\frac{c_j}{\rho A}\int_0^th_j(t-\tau)F^i(\tau)d\tau
\label{MC y}
\end{equation}

As last step, after the evaluation of the $n\times N$ response processes $Y_j^i(t)$ with $i=1,2\dots,N$, and $j=1,2,\dots,n$, the stochastic displacement process samples $v^i(z,t)$ can be used to evaluate the statistics numerically. 



\subsection{Frequency domain approach}

Steady-state analysis and characterization of stationary responses can be driven in analytical way. 
Specifically, we make the truncated Fourier transform of the stochastic differential equation~(\ref{eq forc sto}). That is,
\begin{equation}
 \hat Y_j(\omega,\mathrm{T})\left[  -\omega^2 +\frac{ \eta}{\rho A} \mathrm{i}\omega+\frac{k_{\lambda,j}}{\rho A}\right]=\frac{c_j}{\rho A} \hat F(\omega,\mathrm{T})
\label{eq forc freq}
\end{equation}
which transforms the differential equation in time domain to an algebraic equation in frequency domain. From Eq.~(\ref{eq forc freq}) we get the solution $ \hat Y_j(\omega,\mathrm{T})$ as
\begin{equation}
\begin{split}
 \hat Y_j(\omega,\mathrm{T})&=\frac{1}{  -\omega^2 \hat +\frac{ \eta}{\rho A} \mathrm{i}\omega+\frac{k_{\lambda,j}}{\rho A}} \frac{c_j}{\rho A}\hat F(\omega,\mathrm{T})  \\
 &=H_j(\omega)\frac{c_j}{\rho A}\hat F(\omega,\mathrm{T})
\label{eq forc freq2}
\end{split}
\end{equation}
where $H_j(\omega)$ is the transfer function of the $j$-th modal oscillator.
Once the response $ \hat Y_j(\omega)$ in frequency domain is known, we can place it in the expression of the cross PSD of the response processes $Y_j(t)$ and $Y_i(t)$. That is,
\begin{equation}
\begin{split}
S_{Y_jY_i}(\omega)&=\lim_{\mathrm{T}\to\infty}\frac{1}{2\pi\mathrm{T}}\mathbb{E}\left[\hat Y_j^*(\omega,\mathrm{T})\hat Y_i(\omega,\mathrm{T})\right]\\
&=\frac{c_j}{\rho A}H_j^*(\omega)\lim_{\mathrm{T}\to\infty}\frac{1}{2\pi\mathrm{T}}\mathbb{E}\left[\hat F^*(\omega,\mathrm{T})\hat F(\omega,\mathrm{T})\right]\frac{c_i}{\rho A}H_i(\omega)\\
&=\frac{c_jc_i}{\left(\rho A\right)^2}H_j^*(\omega)H_i(\omega)S_F(\omega)
\end{split}
\label{cross psd}
\end{equation}

Finally, by using Eq.~(\ref{cross psd}), it is possible to evaluate the analytical form of the PSD of beam displacements as follows
\begin{equation}
\begin{split}
S_v(z,\omega)&=\sum_{j=1}^\infty\sum_{i=1}^\infty\phi_j(z)\phi_i(z)\lim_{\mathrm{T}\to\infty}\frac{1}{2\pi \mathrm{T}}\mathbb{E}\left[\hat Y_j^*(\omega,\mathrm{T})\hat Y_i(\omega,\mathrm{T})\right]  \\
&=\sum_{j=1}^\infty\sum_{i=1}^\infty\phi_j(z)\phi_i(z)S_{Y_jY_i}(\omega) \\
&=\frac{S_F(\omega)}{(\rho A)^2}\sum_{j=1}^\infty\sum_{i=1}^\infty\phi_j(z)\phi_i(z){c_jc_i}H_j^*(\omega)H_i(\omega)
\label{psd resp}
\end{split}
\end{equation}

The PSD in Eq.~(\ref{psd resp}) allows for evaluating the stationary variance of transverse displacements 
$\sigma^2_v(z)$. That is,
\begin{equation}
\sigma^2_v(z)=\int_{-\infty}^\infty S_v(z,\omega)d\omega  =\frac{1}{(\rho A)^2}\sum_{j=1}^\infty\sum_{i=1}^\infty\phi_j(z)\phi_i(z)c_jc_i\int_{-\infty}^\infty H_j^*(\omega)S_F(\omega)H_i(\omega)d\omega
\label{variance}
\end{equation}

\subsection{Nonlocal beam under Gaussian white noise}
In this section we consider the case in which the input process is a Gaussian white noise with zero mean denoted by $W(t)$ and characterized by a constant PSD and a Dirac delta as CF. That is,
\begin{equation}
S_W(\omega)=S_0,\;\;\;R_W(\tau)=2\pi S_0\delta(\tau)
\label{psd w}
\end{equation}
With this assumption the present study does not lose generality inasmuch several real excitation process can be modeled as summation of modulated white noises. 

\break
Moreover, from this considered case analytical solutions in terms of statistics of the response can be obtained and some useful results about the time and frequency domain analysis can be drawn.
Under the assumptions in Eq.~(\ref{psd w}), Eq.~(\ref{psd resp}) and Eq.~(\ref{variance}) provide the characterization of the stochastic output process in terms of PSD and stationary variance. Specifically,
\begin{equation}
S_v(z,\omega)=\frac{S_0}{(\rho A)^2}\sum_{j=1}^\infty\sum_{i=1}^\infty\phi_j(z)\phi_i(z){c_jc_i}H_j^*(\omega)H_i(\omega)
\label{psd resp w}
\end{equation}
and 
\begin{equation}
\sigma^2_v(z)=\frac{S_0}{(\rho A)^2}\sum_{j=1}^\infty\sum_{i=1}^\infty\phi_j(z)\phi_i(z)c_jc_i\int_{-\infty}^\infty H_j^*(\omega)H_i(\omega)d\omega
\label{variance w}
\end{equation}
However, the quantities in Eqs.~(\ref{psd resp w}) and (\ref{variance w}) provide a characterization of the displacement just at steady-state. In order to provide a complete stochastic characterization of the response it is needed to evaluate the CF in Eq.~(\ref{comp modal}) and the time-dependent non-stationary variance in Eq.~(\ref{comp modal var}). 
In this regard, exploiting the properties  of the stochastic input, the CF of the displacement is given by 
\begin{equation}
R_v(z,t_1,t_2)=\sum_{j=1}^\infty\sum_{i=1}^\infty\phi_j(z)R_{Y_jY_i}(t_1,t_2)\phi_i(z)
\label{comp modal w}
\end{equation}
where each term in the summations can be evaluated in closed form. That is,
\begin{equation}
\begin{split}
R_{Y_jY_i}(t_1,t_2)&=\frac{c_jc_i 2\pi S_0}{(\rho A)^2}\int_0^{t_1}\int_0^{t_2} h_j(t_1-\tau_1)h_i(t_2-\tau_2)\delta(\tau_2-\tau_1)d\tau_1d\tau_2      \\
	&=\frac{c_jc_i 2\pi S_0}{(\rho A)^2}\int_0^{t_1} h_j(t_1-\tau_1)h_i(t_2-\tau_1)d\tau_1  
\end{split}
\label{comp modal w jk}
\end{equation}

From Eq.~(\ref{comp modal w}) the time-dependent variance of the displacement is
\begin{equation}
\begin{split}
\sigma_v^2(z,t)&=R_v(z,t,t)=\sum_{j=1}^\infty\sum_{i=1}^\infty\phi_j(z)\phi_i(z)R_{Y_jY_i}(t,t)\\
	&=\frac{2\pi S_0}{(\rho A)^2}\sum_{j=1}^\infty\sum_{i=1}^\infty c_jc_i\phi_j(z)\phi_i(z)\int_0^{t} h_j(t-\tau)h_i(t-\tau)d\tau  \\
\end{split}
\label{comp modal var w jk}
\end{equation}

Note that both the expressions in Eq.~(\ref{comp modal w jk}) and in Eq.~(\ref{comp modal var w jk}) can be evaluated in analytical form. 
%
%

\section{Numerical simulation and parametric study}\label{sect5}

This section is devoted to the stationary and non-stationary analysis of the stochastic response of a nonlocal damped Bernoulli-Euler beam forced by a random load. Such numerical analysis aims to study the influence of the nonlocal parameter $\lambda$ in the response in terms of CF, PSD and stationary and non-stationary variance.  

\begin{figure}[!hbt]
    \centering
    $\;\;\;\;\;\;\;\;\;\;\;$
    \begin{subfigure}[b]{0.75\textwidth}
        \includegraphics[width=\textwidth]{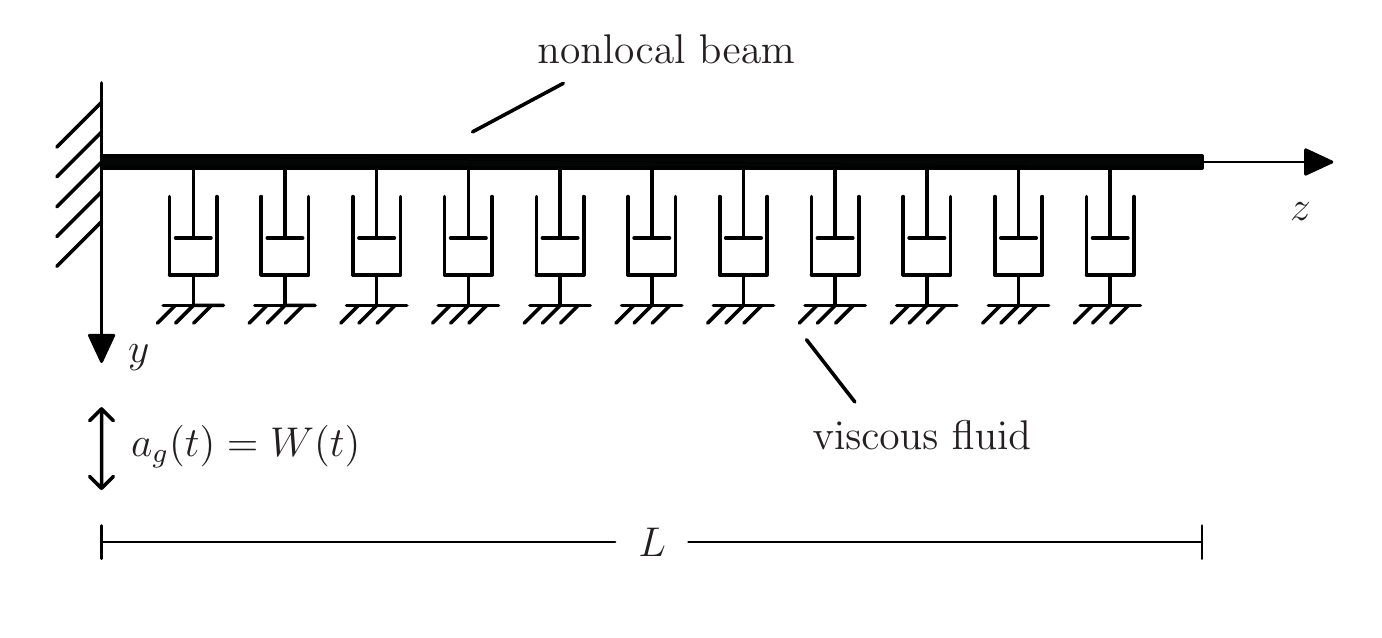}
        \caption{Layout of the cantilever micro-beam}
        \label{cant nanob}
    \end{subfigure}\\
        \vspace{5mm}
    \begin{subfigure}[b]{0.65\textwidth}
        \includegraphics[width=\textwidth]{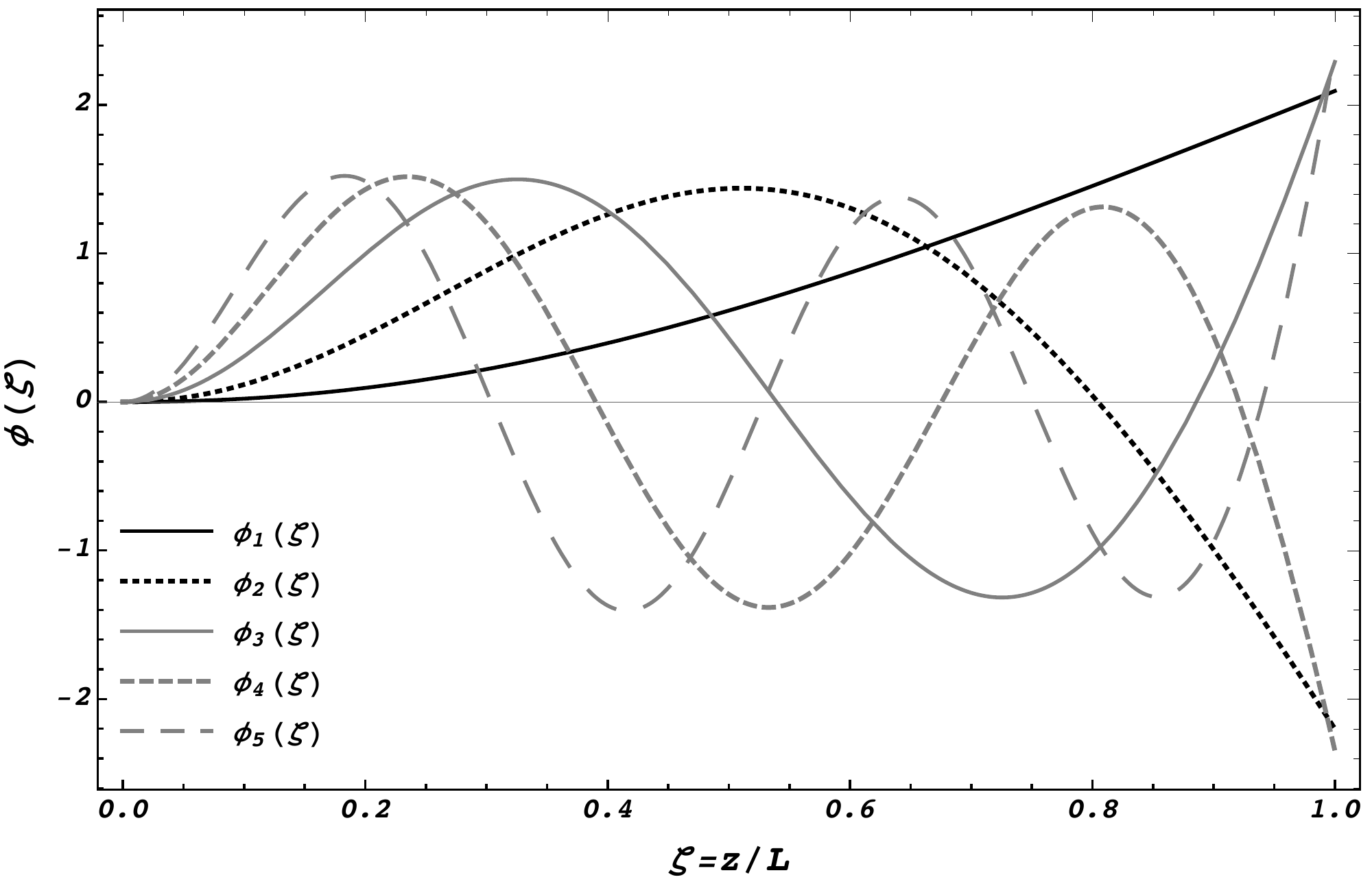}
        \caption{First five eigenfunctions for $\lambda=0.2$}
        \label{cant eigen}
    \end{subfigure}
    \caption{Cantilever micro-beam forced by ground motion acceleration}
    \label{cantilever beam}
\end{figure}

We consider a micro-beam of length $L=300$~$\mu$m, rectangular cross section with width $b=30$~$\mu$m and thickness $h=25$~$\mu$m, made of epoxy characterized by density $\rho=1.20$~g/cm$^3$ and elastic modulus $E=2.20$~GPa \cite{Mems,Ashby1}.
Damping effects due to surrounding environment are modeled by the following value of viscosity $\eta=2.00$~cP describing a wide variety of viscous fluids of technical interest \cite{Ashby1,Ashby2}.
The considered micro-beam is constrained as a cantilever beam and is forced by a ground motion acceleration as depicted in \figurename~\ref{cant nanob}. Without loss of generality, such imposed acceleration on the basis $z=0$ is a Gaussian white noise
\begin{equation}
\ddot v(0,t)=a_g(t)=W(t)
\end{equation}
where 
the white noise $W(t)$ has zero-mean and constant PSD $S_0=10^3$ N$^2$s, and $v(z,t)$ denotes the relative displacement with respect to the basis. In this case the partial differential equation (\ref{equilibrio EB}) becomes
\begin{equation}
\frac{\partial^4v(z,t)}{\partial z^4}-(\lambda L)^2\frac{\partial^6v(z,t)}{\partial z^6}+\frac{\eta}{EI} \frac{\partial v(z,t)}{\partial t}+\frac{\rho A}{EI} \left[\frac{\partial^2 v(z,t)}{\partial t^2}+a_g(t)\right]=0,\;\;\;\;\;0<z<L
\label{equilibrio EB cant}
\end{equation}
and the BCs $\forall t$ are 
\begin{equation}
\left\{\begin{split}
&v(0,t)=0, 		&\;\;\;\;				&M(L,t)=0	,\\
&\varphi(0,t)=0	,	&\;\;\;					&T(L,t)=0
\end{split}\right.
\label{bc cant}
\end{equation}
Let us recall that, by virtue Eq.\eqref{eq3.01} and Eq.\eqref{momento}, bending moment $\,M\,$ and shear force $\,T\,$ fields can be expressed in terms of elastic curvature and its derivatives as
\begin{subequations}
\begin{equation}
M(z,t)=EI\chi(z,t) -EI (\lambda L)^2\frac{\partial^2\chi(z,t)}{\partial z^2}
\label{momento nl}
\end{equation}
\begin{equation}
T(z,t)=EI\frac{\partial\chi(z,t)}{\partial z} -EI (\lambda L)^2\frac{\partial^3\chi(z,t)}{\partial z^3}
\label{taglio nl}
\end{equation}
\end{subequations}

To show the effects of the nonlocal parameters in the response, different cases of Eq.~(\ref{equilibrio EB cant}) are considered. Specifically, we select three values of the nonlocal parameter $\lambda$, i.e., $\lambda=\left\{0.1,\, 0.2,\,0.3\right\}$. 

The solution in terms of displacement function is obtained with the aid of Eq.~(\ref{approx}), where each eigenfunction is obtained by solving the differential problem in Eq.~(\ref{eigenfunction}) with the two constitutive BCs in Eq.~(\ref{2bc}) and the four BCs in Eq.~(\ref{bc cant}). The latter BCs in terms of eigenfunctions are
\begin{equation}
\left\{
\begin{split}
&\phi(0)=0, 		&\;			&\phi^{(2)}(L)-(\lambda L)^2\phi^{(4)}(L)=0	,\\
&\phi^{(1)}(0)=0	,	&\;\;\;				&\phi^{(3)}(L)-(\lambda L)^2\phi^{(5)}(L)=0
\end{split}\right.
\end{equation}

The first five eigenfunctions for $\lambda=0.2$ are shown in \figurename~\ref{cant eigen} and the first five natural frequencies for different values of $\lambda$ are reported in \tablename~\ref{tab1}.

\begin{table}[hbt]
  \begin{center}
    \begin{tabular}{l c c c c c c}
    \hline
       \vspace{-0.35cm}   \\
  $\lambda$	& $ \omega_{0,1}$ 		& $ \omega_{0,2}$ 		& $ \omega_{0,3}$		& $ \omega_{0,4}$		& $ \omega_{0,5}$ \\
     \vspace{-0.35cm}   \\
    \hline
       \vspace{-0.35cm}   \\
    	0.10		& $4.2323\times 10^5$ 		&  $2.8373\times 10^6$	& $8.8780\times 10^6$		&$1.9924\times 10^7$ 	&$3.7820\times 10^7$\\
	 0.15		& $ 4.4551\times 10^5$ 		&  $ 3.1643\times 10^6$	& $ 1.0609\times 10^6$		&$ 2.5278\times 10^7$ 	&$ 5.0090\times 10^7$\\
	0.20	  	& $4.6795\times 10^5$		&  $3.5244\times 10^6$	& $1.2495\times 10^7$		&$3.1001\times 10^7$	&$6.2992\times 10^7$\\
	 0.25		& $ 4.9002\times 10^5$ 		&  $ 3.9038\times 10^6$	& $ 1.4464\times 10^6$		&$ 3.6905\times 10^7$ 	&$ 7.6186\times 10^7$\\
    	0.30		& $5.1192\times 10^5$		&  $4.2951\times 10^6$	& $1.6481\times 10^7$		&$4.2909\times 10^7$	&$8.9539\times 10^7$\\
      \vspace{-0.35cm}   \\
       \hline
    \end{tabular}
  \end{center}
\caption{Natural frequencies in $rad/s$ of cantilever micro-beam for different values of $\lambda$.}
\label{tab1}
\end{table}


Taking Eq.~(\ref{equilibrio EB cant}) into account and with the aid of the definition in Eq.~(\ref{psd resp w}) the PSD of the displacement is
\begin{equation}
S_v(z,\omega)\approx S_0\sum_{j=1}^n\sum_{i=1}^n\phi_j(z)\phi_i(z)c_jc_iH_j^*(\omega)H_i(\omega)
\label{psd resp approx}
\end{equation}
where for the present numerical simulations we assume $n=5$. 

\begin{figure}[!hbt]
    \centering
        \begin{subfigure}[b]{0.65\textwidth}
        \includegraphics[width=\textwidth]{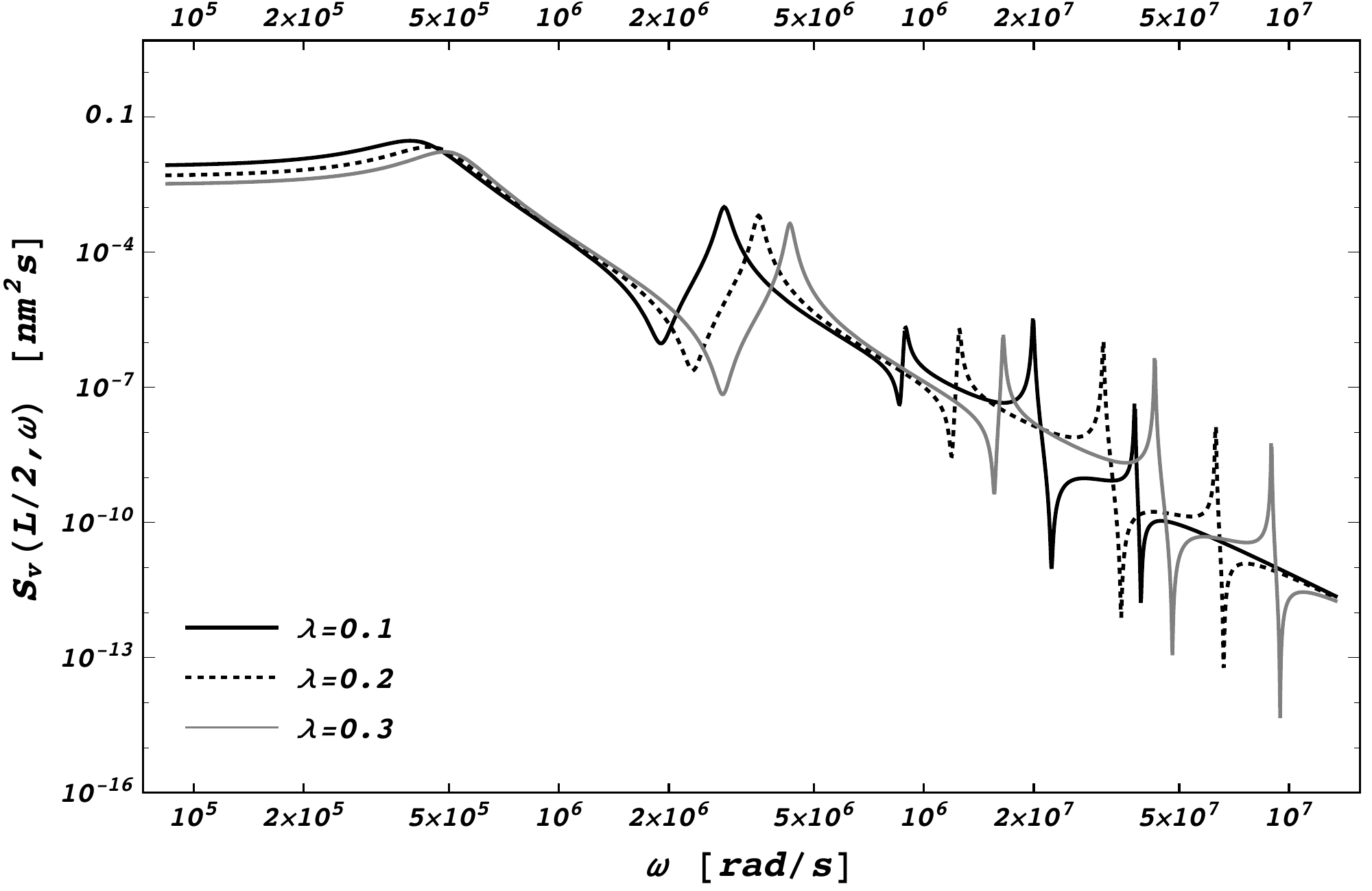}
        \caption{PSD for $z=L/2$}
        \label{PSD mezz}
    \end{subfigure}
    \begin{subfigure}[b]{0.65\textwidth}
        \includegraphics[width=\textwidth]{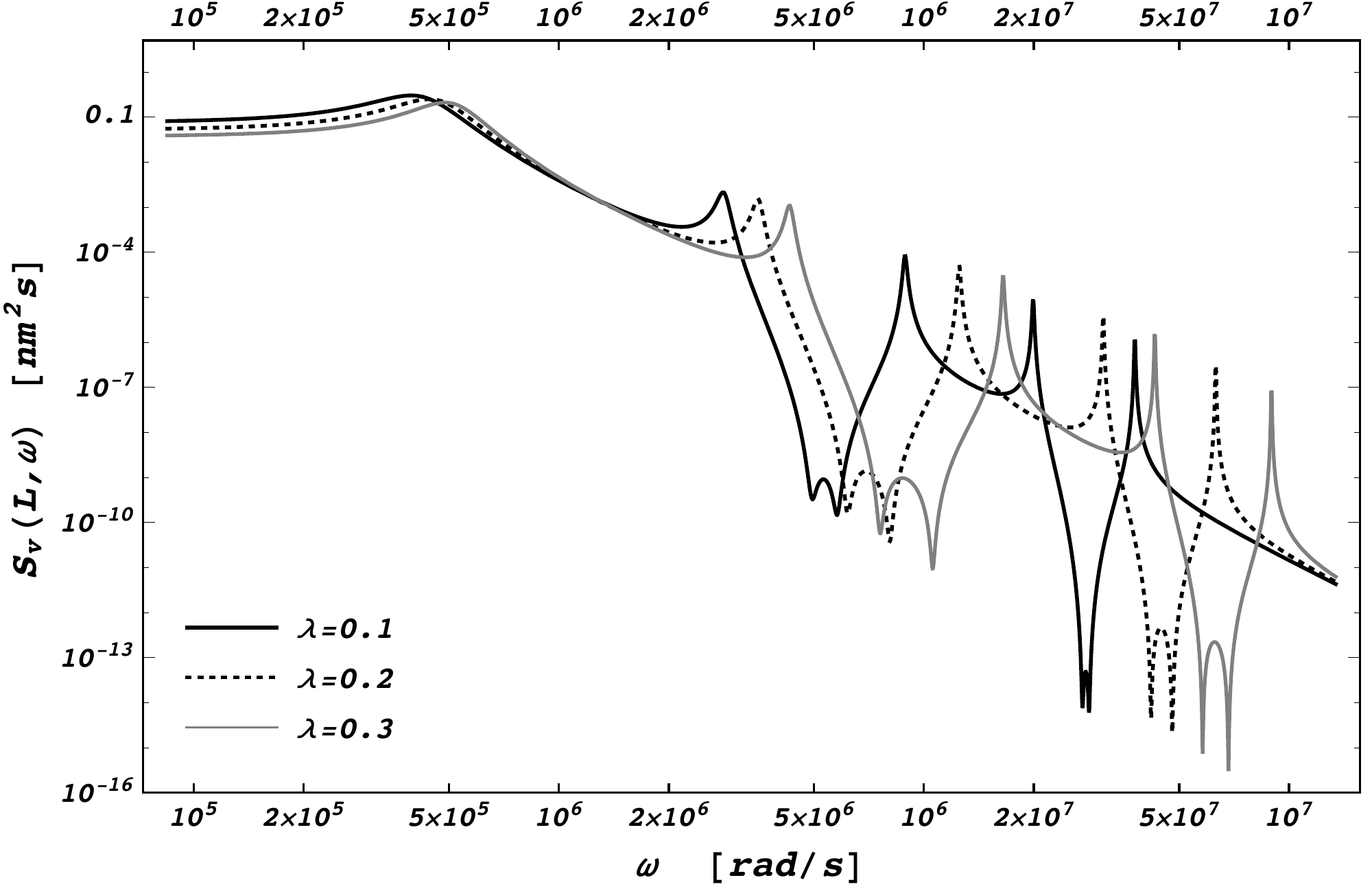}
        \caption{PSD for $z=L$}
        \label{PSD L}
    \end{subfigure}$\;$
    \caption{PSD of a cantilever micro-beam response for different values of $\lambda$}
    \label{PSD cantilever}
\end{figure}

In \figurename~\ref{PSD cantilever} the PSD of the displacements are reported for different values of $\lambda$. Specifically, \figurename~\ref{PSD mezz} shows the PSDs of the mid-point displacements, while in
\figurename~\ref{PSD L} the PSDs at $z=L$ are shown. We can observe that the nonlocal parameter influences both the natural frequencies and the peak amplitudes in the PSD. Specifically, when the nonlocal parameter increases the amplitude of the peaks decrease and their frequencies increases.
From the PSD in Eq.~(\ref{psd resp approx}) and with the aid of Eq.~(\ref{variance}) we can also evaluate the stationary variance of the displacement. Such quantity can be evaluated by the integration of the PSD in Eq.~(\ref{psd resp approx}) as follow
\begin{equation}
\sigma_v^2(z)=\int_{-\infty}^\infty S_v(z,\omega)d\omega
	\approx S_0\sum_{j=1}^n\sum_{i=1}^n\phi_j(z)\phi_i(z)c_jc_i \int_{-\infty}^\infty H_j^*(\omega)H_i(\omega)d\omega
	\label{stat var w gm}
\end{equation}

In \tablename~\ref{tab2} stationary variances at $z=L/2$ and $z=L$ for different values of $\lambda$ are reported. 
\begin{table}[hbt]
  \begin{center}
    \begin{tabular}{l c c c }
    \hline
   \vspace{-0.35cm}   \\

  $\lambda$	&& $\sigma^2_v(L/2)$ 	& $ \sigma^2_v(L)$ 	\\
   \vspace{-0.35cm}   \\
      \hline
    \vspace{-0.35cm}   \\
    	0.10		&& $0.0200$ 			&  $0.1975$	\\
    	0.15		&& $ 0.0165$ 			&  $ 0.1791$	\\
  	0.20	  	&& $0.0144$			&  $0.1645$	\\
	 0.25		&& $ 0.0138$ 			&  $ 0.1548$	\\
    	0.30		&& $0.0115$			&  $0.1416$	\\
      \vspace{-0.35cm}   \\
       \hline
    \end{tabular}
  \end{center}
\caption{Displacement variances in $\mu$m$^2$ at $z=L/2$ and $z=L$, for different values of $\lambda$.}
\label{tab2}
\end{table}
Taking into account the values in this table we can state that the nonlocal parameter also influences stationary variances. Specifically, when the nonlocal parameter $ \lambda $ grows up then the displacement stationary variances decrease.

The PSD in Eq.~(\ref{psd resp approx}) and the stationary variance in Eq.~(\ref{stat var w gm}) provide a steady-state characterization of the displacement process due to a stochastic ground motion acceleration. However, taking   Eq.~(\ref{equilibrio EB cant}) into account, a characterization of the non-stationary response can be pursued by using Eq.~(\ref{comp modal w}) to evaluate the CF of the process $v(z,t)$. That is,
\begin{equation}
R_v(z,t_1,t_2)=2\pi S_0\sum_{j=1}^\infty\sum_{i=1}^\infty\phi_j(z)\phi_i(z){c_jc_i }\int_0^{t_1} h_j(t_1-\tau_1)h_i(t_2-\tau_1)d\tau_1  
\label{comp modal w jk gm}
\end{equation}
By virtue of the definition in Eq.~(\ref{comp modal var w jk}) the time-dependent variance is 
\begin{equation}
\begin{split}
\sigma_v^2(z,t)&=R_v(z,t,t)=\sum_{j=1}^\infty\sum_{i=1}^\infty\phi_j(z)\phi_i(z)R_{Y_jY_i}(t,t)\\
	&={2\pi S_0}\sum_{j=1}^\infty\sum_{i=1}^\infty c_jc_i\phi_j(z)\phi_i(z)\int_0^{t} h_j(t-\tau)h_i(t-\tau)d\tau  \\
\end{split}
\label{comp modal var w jk gm}
\end{equation}
From Eq.~(\ref{comp modal var w jk gm}) it is possible to obtain the stationary variance in Eq.~(\ref{stat var w gm}) by performing the following limit
\begin{equation}
\sigma_v^2(z)=\lim_{t\to\infty} \sigma_v^2(z,t)
\end{equation}
while the stationary CF is given by
\begin{equation}
R_v(z,\tau)=\lim_{t\to\infty} R_v(z,t,t+\tau)
\end{equation}
the latter equation is related to the PSD in Eq.~(\ref{psd resp approx}) by Fourier transform (Wiener-Khinchin theorem). 
\begin{figure}[h]
    \centering
        \begin{subfigure}[b]{0.65\textwidth}
        \includegraphics[width=\textwidth]{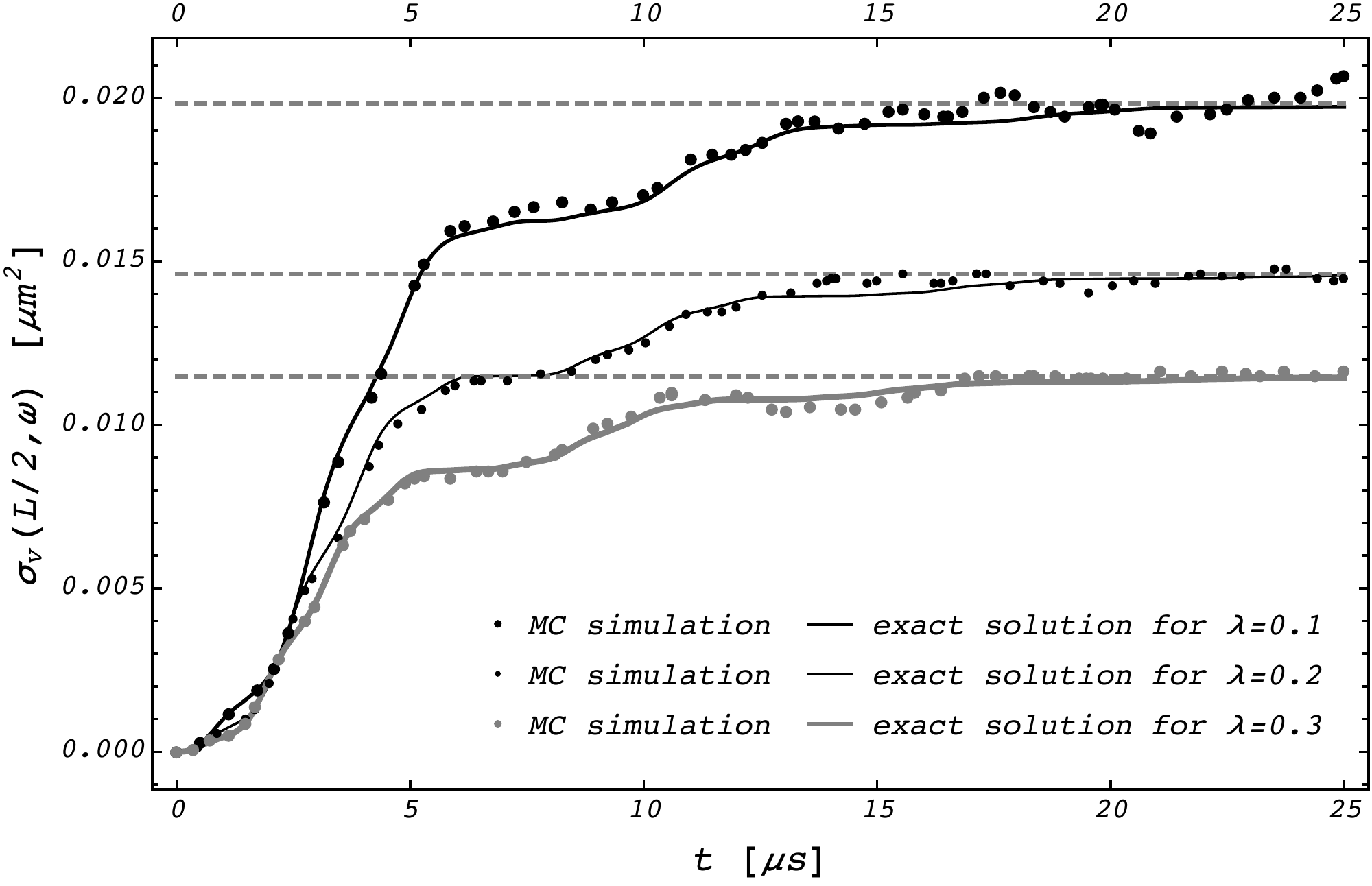}
        \caption{Displacement variance at $z=L/2$}
        \label{var mezz}
    \end{subfigure}
    \begin{subfigure}[b]{0.65\textwidth}
        \includegraphics[width=\textwidth]{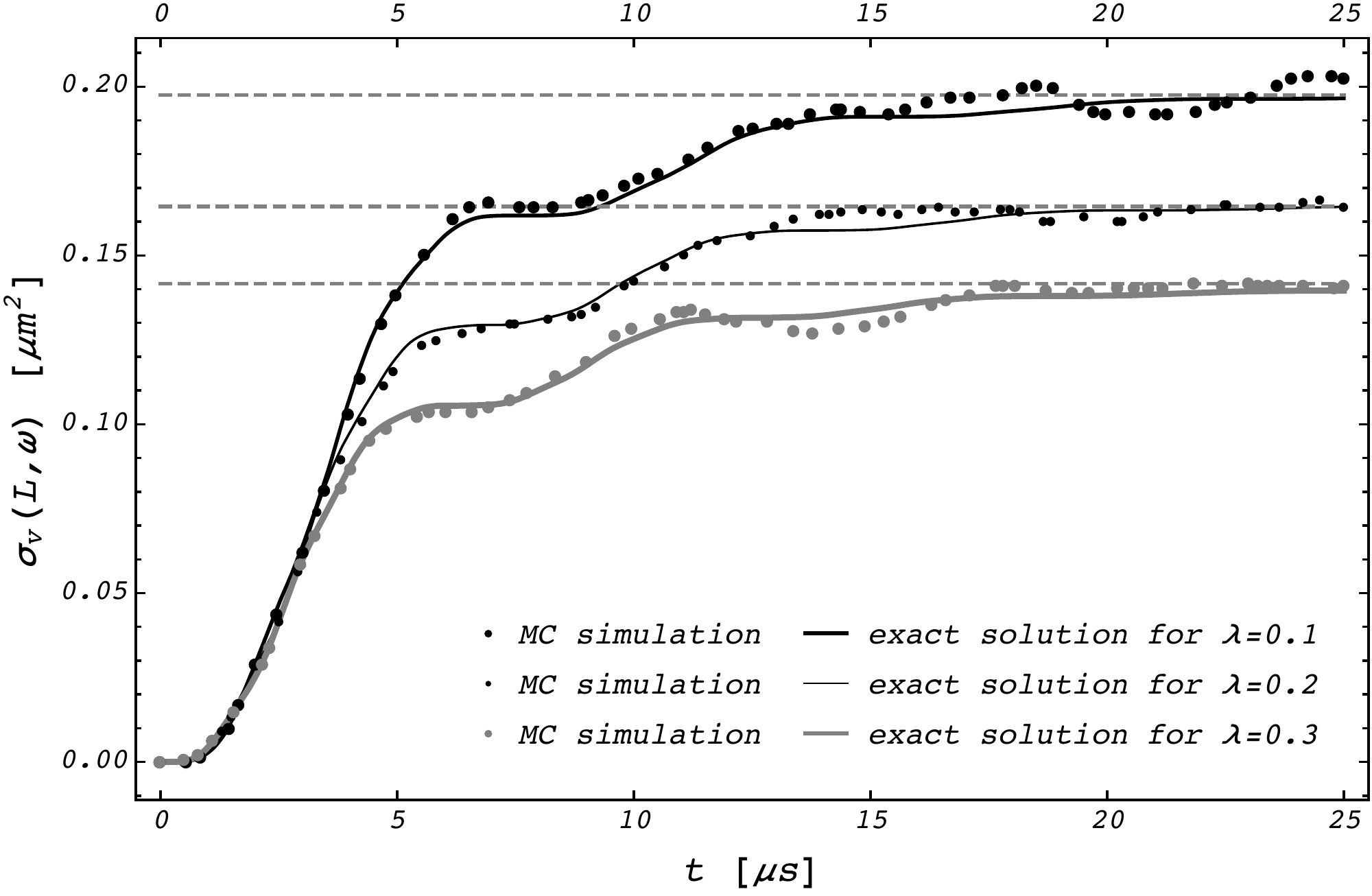}
        \caption{Displacement variance at $z=L$}
        \label{var L}
    \end{subfigure}$\;$
    \caption{Exact stationary (dashed line) and non-stationary (continuous line) displacement variance in contrast with those obtained by MC simulations}
    \label{Var cantilever}
\end{figure}
\figurename~\ref{Var cantilever} shows the stationary and non-stationary displacement variances for different values of the nonlocal parameter $\lambda$. Dashed lines represent the stationary variances, whereas continuous lines are the non-stationary ones obtained with the aid of Eq.~(\ref{comp modal var w jk gm}). Such exact variances are compared with the numerical results obtained by means of the MC approach described in Section~\ref{MC}. Specifically, the numerical variances (dotted lines in \figurename~\ref{Var cantilever}) are obtained considering $N=4\times10^3$ samples and by assuming $m=2\times10^3$ and $\Delta\omega=\omega_{0,1}/50$ in Eq.~(\ref{MC F}). From \figurename~\ref{Var cantilever} we can observe that the steady state is reached in all considered case in a few tens of microseconds. This is due to the fact that the involved stiffnesses in such micro-beams are great and the masses are little. Moreover, the nonlocal parameter influences the duration of the transient state, indeed, if $\lambda$ grows up the variance reaches the stationary value more quickly.

\section{Concluding remarks}
\label{sect6}

Random vibrations of damped nonlocal Bernoulli-Euler beams due to stochastic excitation have been investigated in the present research. 
Two specific effects have been accurately analyzed: size and damping phenomena, respectively modeled by stress-driven nonlocal mechanics and external viscous interactions.
A stochastic input for the loading has been assumed to simulate external actions randomness.

A stochastic differential problem in space and time, governing the motion of nonlocal beams under stochastic loading, has been formulated. 

Exact solutions of power spectral density, correlation function and displacement variance have been evaluated 
by differential eigenanalysis. 

From the analytical formulation of stationary and non-stationary responses and with the aid of numerical simulations, it has been highlighted a significant reduction of stationary variances and duration of the transient state in responses
and an increasing of natural frequencies for increasing nonlocal scale parameter.
The predicted smaller-is-stiffer phenomenon, confirmed recently in \cite{FusPisPol}, agrees with most of experimental outcomes associated with inflected small-scale beams \cite{AbazariSensors2015}.

In summary, the nonlocal approach developed to model damped small-scale beams is able to capture size and damping effects and random excitations. 
The methodology provides analytical solutions in terms of statistics of the response and closed-form natural frequencies. 
The contributed results can be exploited for structural design and optimization of smaller and smaller devices used in modern technological applications, such as: sensors, actuators, MEMS/NEMS, resonators.

\section*{Acknowledgments}
Financial supports from the MIUR in the framework of the Projects PRIN 2015 "COAN 5.50.16.01" (code 2015JW9NJT \emph{Advanced mechanical modeling of new materials and structures for the solution of 2020 Horizon challenges}) and
PRIN 2017 (code 2017J4EAYB \emph{Multiscale Innovative Materials and Structures (MIMS)}; University of Naples Federico II Research Unit)
and from the research program ReLUIS 2019 are gratefully acknowledged.


\section*{Conflict of interest}
The authors declare that they have no conflict of interest.

\end{document}